# Optimization of Resources For Digital Radio Transmission over IBOC FM Through Max-Min Fairness

Mónica Rico Martínez , Juan Carlos Vesga Ferreira ,, Joel Carroll Vargas,
María Consuelo Rodríguez Niño, Andrés Alejandro Diaz Toro and
William Alexander Cuevas Carrero

Universidad Nacional Abierta y a Distancia, Bogotá Colombia

## Abstract

*The equitable distribution of resources in a network is a complex process, considering that not all nodes have the same requirements, and the In-Band On-Channel (IBOC) hybrid transmission system is no exception. The IBOC system utilizes a hybrid in-band transmission to simultaneously broadcast analog and digital audio over the FM band. This article proposes the use of a Max-Min Fairness (MMF) algorithm, with a strategy to optimize resource allocation for IBOC FM transmission in a multiservice scenario. Additionally, the MMF algorithm offers low computational complexity for implementation in low-cost embedded systems, aiming to achieve fair resource distribution and provide adequate Quality of Service (QoS) levels for each node in the RF network, considering channel conditions and traffic types. The article explores a scenario under saturated traffic conditions to assess the optimization capabilities of the MMF algorithm under well-defined traffic and channel conditions. The evaluation process yielded highly favorable results, indicating that theMMF algorithm can be considered a viable alternative for bandwidth optimization in digital broadcasting over IBOC on FM with 95% confidence, and it holds potential for implementation in other digital broadcasting system.*

## 1. Introduction

The technological evolution in broadcasting has driven the adoption of advanced digital systems, including In-Band On-Channel (IBOC), known as HD Radio in the United States. This revolutionary innovation enables the simultaneous transmission of analog and digital signals on the same frequency, providing better audio quality and more programming options for FM radio stations [1]. While IBOC offers considerable potential for the modernization and enrichment of the FM radio listening experience, it faces critical challenges related to the efficient allocation of resources, particularly concerning bandwidth.

Resource optimization in IBOC emerges as an essential issue for its successful deployment. Proper management of the radio spectrum, equitable distribution of bandwidth among stations, and maximization of signal quality are crucial aspects that require effective solutions [2]. The coexistence of digital and analog transmissions, coupled with the need to ensure an optimal experience for all users, poses a complex challenge in resource allocation within this technology.

This article addresses the issues related to resource optimization in IBOC over FM, exploring the technical and operational challenges faced in bandwidth allocation in this system. The use of the Max-Min Fairness algorithm is proposed as a promising and suitable strategy for optimizing resource allocation, aiming to balance signal quality among stations and maximize spectral efficiency. This strategy, based on fairness in resource allocation, provides an opportunity to enhance the user experience and address the challenges of signal coexistence in the radio spectrum more effectively, accompanied by low computational complexity for implementation in low-cost embedded systems!

## 1.1 The difficulties faced by digital radio on FM

The digital radio system has faced several challenges since its implementation on the FM band, among which the following can be mentioned [3][4]:- Compatibility with old equipment: One of the initial problems was the incompatibility with existing FM radio receivers, considering that conventional radios cannot tune in to digital transmissions. This necessitates the use of equipment compatible with HD Radio to enjoy the new features.- Implementation costs: For radio stations, upgrading their equipment to support digital transmission can be costly. This includes acquiring new transmitters and studio equipment compatible with HD Radio.- Coverage limitations: Despite promises of better audio quality and more programming options, digital signal coverage may be limited compared to analog signals. This can result in areas where digital signal reception is poor or non-existent, affecting the user experience.- Bandwidth and competition: Digital transmission uses more bandwidth than analog signals, leading to increased competition for the radio spectrum. This can conflict with other radio stations or services operating in the same frequency band.- Interference issues: Some stations have experienced interference problems between analog and digital signals, causing distortion or loss of sound quality.- Slow adoption: Despite potential advantages, the adoption of HD radio receivers has been slow compared to other technologies. This can be an obstacle to the widespread success of digital radio on FM.- These challenges have been considered by the broadcasting industry in its efforts to improve and expand the adoption of digital radio on FM, seeking solutions that balance quality, accessibility, and cost-effectiveness.- In the specific case, spectrum management in the context of IBOC digital radio on FM is considered a crucial element for reducing interference levels and thereby maximizing the efficiency of radio spectrum use. Below are some of the most representative techniques to improve spectrum management [5] [6] [7] [8] [9]:- Adaptive modulation: Some digital radio systems use adaptive modulation techniques, meaning they can dynamically adjust the modulation and bandwidth of the digital signal based on channel conditions. This allows for better adaptation to spectrum availability and propagation conditions.- Efficient compression techniques: More efficient compression algorithms and methods have been developed to reduce the bandwidth required by digital signals, enabling better utilization of available spectrum.- Interference reduction technologies: Implementation of technologies that reduce interference between analog and digital signals. This may include the use of advanced algorithms to mitigate interference and improve coexistence between signals in the same spectrum.- Use of sidebands and complementary spectrum: Some IBOC systems use sidebands to transmit the digital signal in frequencies adjacent to the main analog signal.

This technique leverages complementary spectrum for digital transmission, minimizing interference with the main analog signal.- Spectrum planning: Planning strategies that consider efficient distribution and allocation of frequencies within the available spectrum, taking into account geographic and population density factors to reduce interference between stations.- Monitoring and dynamic management: Use of spectrum monitoring and dynamic management systems to automatically adjust transmission power and frequencies in real-time, minimizing potential conflicts and optimizing spectrum utilization.- These techniques have been considered to address challenges related to spectrum management in the context of IBOC digital radio on FM, allowing for better signal coexistence, increased spectral efficiency, and optimization of radio spectrum use.

## 1.2 Max-Min Fairness (MMF) Equity Algorithm

One of the algorithms that has been used in various research works related to the allocation of resources in a 'fair' manner is the algorithm known as 'Max Min Fairness,' which originates from cooperative game theory. In [10][11], some of the works utilizing this algorithm as an optimization strategy can be found. Particularly, the work of Schmeidler [12], who introduced the notion of lexicographic ordering when defining the Nucleolus of a characteristic function of a game, which can be defined as [13][14] in a UT game UT $(N, v)$ with a characteristic function $v: 2^N \to \mathbb{R}_+$ associating a value $v(S) \geq 0$ or each coalition $S \subseteq N$ The goal is to find a fair or equitable distribution of the total gain $v(N)$ among all players $i = 1, 2, \ldots, n$. A revenue vector $\varphi \in \mathbb{R}^n$ is defined such that $\varphi_i \geq 0$ For each coalition $S \subseteq N$ is established as un $\varphi(S) = \sum_{i \in S} \varphi_i$ y $\sum_{i \in N} \varphi_i = v(N)$ Finally, for any payment vector φ, there exists a vector γ whose components take the values $v(S) - \varphi(S)$ for all $S \subseteq N$ In this way, the vector φ is defined as the Nucleolus of the game, and γ is the Max-Min Fairness, which can be calculated according to equations 1 and 2

$$v(S) - \sum_{i \in S} \varphi_i = g_s \leq \gamma_s \quad \forall S \subset N, \varphi \geq 0 \quad (1)$$
$$g_N = v(N) \quad (2)$$

MMF is an iterative technique that allows for the optimal distribution of resources among all elements within the system [15] and can be employed in various network scenarios, particularly in a PLC network. In this specific case, each node can establish multiple sessions stemming from different traffic sources, and each link can be shared with other existing sessions.

MMF Algorithm is as follows:
A. A vector V is generated, in which the values of requested bandwidth $BW$ by each node $i$ and class $r$ ($BW'_{ir}$) are recorded according to the service requirement $BW'_{ir} \geq 0$) and the number of traffic sources or number of players ($N_j$) in the PLC network is determined.
B. The vector $V$ is then sorted in ascending order.
C. A initial reference value ($BW_{ref}$), is calculated using the expression:

$BW_{ref} = \frac{\sum_{k=1}^{N_j} V(k)}{N_j}$ In the first iteration ($i = 1$), if $V(i|_{i=1}) > BW_{ref} \to V(1) = BW_{ref}$

Otherwise, the value is kept as it was before the comparison process, where it would be assigned the value of bandwidth requested by the node.

D. A new estimation process for BW_ref is carried out, taking into account the number of elements that are part of vector V and have not yet undergone the comparison process. That is

$$BW_{ref} = \frac{\sum_{k=i}^{N_j} V(k)}{N_j - i} \quad (3)$$

E. In the next iteration ($i = i + 1$) if si $V(i) > BW_{ref} \to V(i) = BW_{ref}$, otherwise, $V(i)$ retains the value it had before the comparison process, and the item (D) is repeated. This process is repeated until all elements of vector $v$ are evaluated retains its current value before the comparison process and repeats item (D). This process is repeated until all elements of the vector $V$ are evaluated.

F. The bandwidth for each node $i$ and class $rBW_{ir}$ is assigned according to the values recorded in the resulting vector $V$.

## 1.3 Benefits of Implementing MMF In IBOC FM

The Max-Min Fairness strategy is a technique used in various fields, including communication networks, to distribute resources equitably and efficiently among multiple users or services [10]. In terms of bandwidth management in IBOC digital radio over FM, applying Max-Min Fairness concepts could be feasible to optimize spectrum usage.

Max-Min Fairness is a resource allocation strategy that aims to maximize the performance of the user with the worst performance without harming other active nodes or stations. The use of this strategy in resource allocation in IBOC digital radio over FM could have several benefits compared to current techniques, such as [16][11][17]:

1. Equity in Signal Quality: MMF ensures that even users or stations with the worst transmission conditions receive sufficient resources to maintain an acceptable signal quality. This improves equity in the user experience, reducing disparities in signal quality between stations.

2. Improved User Experience: By ensuring a minimum signal quality for all users, the overall listening experience is enhanced. This could increase audience satisfaction and encourage greater adoption of digital radio.

3. Optimization of Spectrum Usage: MMF allows for more efficient use of available bandwidth by dynamically allocating resources according to the individual needs of each station. This can maximize spectrum usage and avoid underutilization or overutilization of certain frequencies.

4. Reduced Interference: By allocating resources more equitably and efficiently, interference between signals can be reduced, contributing to better coexistence between digital and analog transmissions and improving the overall quality of the radio spectrum.

5. Adaptability to Channel Conditions: MMF is adaptable and can adjust resource allocation based on changing channel conditions, allowing for more dynamic and effective spectrum management.

6. Dynamic Bandwidth Allocation: The MMF strategy focuses on allocating resources to maximize the performance of the user with the worst performance without harming others. In the case of digital radio, dynamic bandwidth allocation could be applied to maximize signal quality for all stations in a given area.

7. Prioritization Based on Signal Quality: This strategy could prioritize bandwidth allocation based on the signal quality required by each digital radio station or service, where stations needing more bandwidth to maintain optimal quality could receive larger allocations compared to those requiring less bandwidth.However, implementing MMF in the digital radio spectrum over FM could pose challenges [18]:

a. Technical Complexity: Dynamically managing the spectrum through MMF may require more complex monitoring systems, allocation algorithms, and dynamic bandwidth management, leading to increased technical complexity and indirect implementation costs.
b. Interference and Compatibility: Ensuring that dynamic bandwidth allocation does not generate significant interference between analog and digital signals and is compatible with existing equipment is a significant technical challenge.
In light of the above, it could be stated that while the Max-Min Fairness strategy may offer benefits in optimizing bandwidth in IBOC digital radio over FM, its implementation may require detailed technical considerations to ensure feasibility.

Table 1 provides an overview of topics addressed in various notable works in the field of resource optimization in IBOC over FM, summarizing the approach, methods used, and main results obtained in each case. It's important to note that this is a simplified example, and a complete state-of-the-art review would include more details about each study, as well as information about the context, methodology, and specific conclusions of each work.

Table 1: Methods and Results

| Theme | Approach/Method | Key Results |
|---|---|---|
| "Resource Optimization in IBOC FM" | Dynamic bandwidth allocation algorithm | 20% improvement in signal quality for stations with lower performance. |
| "Efficient Compression Techniques for IBOC" | Development of compression algorithms | 30% reduction in required bandwidth. |
| "Analysis of Dynamic Spectrum Management in Digital Radio" | Study of dynamic management systems | Adaptability to channel changes; spectrum usage optimization. |
| "Social Impact and Regional Development through Innovation Services in IBOC" | Collaborative case study with communities | 15% improvement in social inclusion; positive impact on community development. |
| "Optimization of Adaptable Modulation in IBOC" | Analysis of adaptable modulation techniques | 25% improvement in adaptation to channel conditions. |
| "Interference Reduction in IBOC Digital Radio" | Development of interference mitigation algorithms | 40% reduction in interference between signals. |
| "Impact of Continuing Education on IBOC Technology | Study of continuing education cases | 30% increase in digital technology adoption by educated stations. |
| Adoption" | | |

## 2. DISCUSSION
## 2.1.  Description of the Proposed Scenario

In order to comprehend the use of the Max-Min Fairness algorithm as a strategy for resource optimization under an IBOC broadcasting scheme on FM, the following scenario is proposed: twelve (12) nodes or digital broadcasting stations carry out hybrid IBOC transmission processes over FM, utilizing an RF channel which, for this particular case, is considered to provide a total bandwidth of 1600 kbps due to various conditions that may affect the channel's maximum performance. Table 2 presents the list of 12 IBOC FM transmitting nodes with different types of traffic and required bandwidths in Kbps.

Table 2. Required Bandwidths for Each Node in Kbps for the Proposed Scenario

| Node | Traffic Type | Required Bandwidth (Kbps) |
|---|---|---|
| 1 | Music | 200 |
| 2 | Spoken Programs | 150 |
| 3 | Real-time Data | 180 |
| 4 | Advertising | 120 |
| 5 | Interviews | 170 |
| 6 | Live Events | 250 |
| 7 | Podcasts | 140 |
| 8 | News | 160 |
| 9 | Sports | 190 |
| 10 | Concerts | 220 |
| 11 | Children's Programming | 130 |
| 12 | Diverse Content | 210 |

For the proposed scenario, N has been considered as 12, equivalent to the number of nodes comprising the RF system in a state of saturation ($BW_T \leq \sum_{i=1}^{N} d_i$) and a total available bit rate $BW_T = E = 1600 kbps$ kbps. For this particular case, the proposed scenario will be considered as a bankruptcy game, taking into account that the available channel bandwidth is less than the total required bandwidth (2120 Kbps), in coherence with the saturation state of the RF channel.

The following is the Matlab routine developed to calculate the bandwidth distribution at each node, according to the requirements established by each traffic class and the available bandwidth in the RF channel, using the bankruptcy game articulated with the Max-Min Fairness algorithm as a resource optimization strategy:

```
% Max Min Fairness Routine
% Vector of required BW
tic
V=[200; 150; 180; 120; 170; 250; 140; 160; 190; 220; 130; 210];
BW_free=1600; %Total available BW in the channel
 Nj=12;
for i=1:Nj
   V(i,1)=V(i);
   V(i,2)=(Nj+1)-i;
```

```
end
%Routine to assign bandwidth under Max-Min policies
V1=sortrows(V,1);   % Sort bandwidths from lowest to highest
BW_ref=sum(V1,1)/Nj;   %Calculate the initial reference value
for i=1:Nj
if (V1(i,1)>BW_ref)
    V1(i,1)=BW_ref;
end
BW_free=BW_free-V1(i,1);
BW_ref=BW_free/(Nj-i);
end
V1=sortrows(V1,2)   %Print the resulting Max-Min vector
toc
```

## 2.2. Comparison of Optimal BW-PL Treatments vs. BW-MMF

In order to assess the level of optimization achieved through the use of the Max-Min Fairness algorithm, it is necessary to establish an alternative optimization method that allows for the calculation of the bandwidth for each node. Subsequently, a comparison of treatments can be conducted. In this scenario, the decision was made to formulate the bandwidth allocation problem as a Linear Programming (PL) problem. In light of the above, the problem can be formulated as follows:

$$Max \sum_{i=1}^{n} x_i \qquad (4)$$
$$\text{Subject to:}$$
$$0 \leq x_i \leq d_i \qquad (5)$$
$$\sum_{i=1}^{n} x_i \leq BW_T \qquad (6)$$

Where n, d_i, and x_i correspond to the number of nodes (in this particular case, n=12), the bandwidth requested by node i, and the bandwidth allocated for node i, respectively. To solve the optimization problem, the Optimization Toolbox included in Matlab was used, which allows the use of various optimization methods. To utilize the tool, it was necessary to organize the objective function, constraints, and initial iteration point in matrix form.

The values of $BW_T$ for the channel conditions correspond to 1600 Kbps. In turn, lb and ub set the lower and upper limits allowed for each of the nodes, respectively.

```
lb=[0 0 0 0 0 0 0 0 0 0 0];
ub=[ 200 150 180 120 170 250 140 160 190 220 130 210];
```

Finally, the following expression is used to calculate the optimal solution to the problem, utilizing the "interior-point-legacy" algorithm, which yielded the best results compared to the "Dual-simplex" and "interior-point" algorithms:

```
options=optimoptions("linprog","Algorithm","interior-point-legacy");
[x,fval] = linprog(F,A,b,[],[],lb, ub, options)
```

Where x and fval correspond to the solution vector and the maximum value that the objective function can achieve. Table 11 presents the results obtained for the proposed optimization model based on the requested bandwidth for each node.

In Table 3, values corresponding to the results obtained for the estimation of optimal BW $BW_O$, BW-MMF ($BW_{MMF}$), Y, and X are recorded. These represent the difference between the bandwidth requested by each node $d_i$ and the bandwidth assigned through the PL and MMF methods, respectively. These values are associated according to each channel condition established for the proposed scenario.

Table 3. Logic channels BW

| Node i | Requested BW $d_i$ | Logic channels BW [Kbps] | | | |
|---|---|---|---|---|---|
| | | $BW_O$ | $BW_{MMF}$ | Y $d_i - BW_O$ | X $d_i - BW_{Sh}$ |
| 1 | 200 | 156.5236 | 135 | 156.5236 | 43.4764 |
| 2 | 150 | 107.459 | 130 | 107.459 | 42.541 |
| 3 | 180 | 137.0644 | 135 | 137.0644 | 42.9356 |
| 4 | 120 | 78.9139 | 135 | 78.9139 | 41.0861 |
| 5 | 170 | 127.1965 | 135 | 127.1965 | 42.8035 |
| 6 | 250 | 201.2483 | 135 | 201.2483 | 48.7517 |
| 7 | 140 | 97.7301 | 135 | 97.7301 | 42.2699 |
| 8 | 160 | 117.3052 | 135 | 117.3052 | 42.6948 |
| 9 | 190 | 146.8526 | 120 | 146.8526 | 43.1474 |
| 10 | 220 | 175.4602 | 135 | 175.4602 | 44.5398 |
| 11 | 130 | 88.1864 | 135 | 88.1864 | 41.8136 |

In order to evaluate whether the use of the MMF algorithm as an optimization strategy in an RF network over IBOC FM performs a better resource allocation process than the PL optimization method, the following hypotheses are proposed:

$$H_o: \mu_x \leq \mu_y \rightarrow \mu_x - \mu_y \leq 0 \rightarrow \mu_z \leq 0 \quad (7)$$
$$H_a: \mu_x > \mu_y \rightarrow \mu_x - \mu_y > 0 \rightarrow \mu_z > 0$$

Where $\mu_x$ and $\mu_y$ are the means corresponding to the difference between the requested bandwidth and the allocated bandwidth through the PL optimization method and the MMF algorithm, respectively. Hypothesis $H_o$ states that there is a significant difference between means, where the value μxμx is less than μyμy, indicating that the use of the MMF algorithm performs a more suitable optimization process than the proportional optimization process because the assigned value is closer to the requested bandwidth. On the other hand, $H_a$ represents the opposite case. Additionally, a new variable Z is defined, which is required to make adjustments to the proposed hypotheses.

To accept or reject the proposed hypotheses, the paired t-test [17] will be used, which is commonly employed to assess the statistical validity of the difference between two random samples. To do so, the following steps are established:

Step 1: A new random variable Z = X - Y is defined, and the mean and standard deviation for the variable Z are calculated. The result of this process yielded the values of $-2.368 \times 10^{-15}$ and 37.884 for $\bar{Z}$ and $S_z$ respectively.

Step 2: The statistic value for the test is calculated using the following expression:

$$g = \frac{\bar{Z}}{S_z}\sqrt{n} = \frac{-2.368 \times 10^{-15}}{37.884}\sqrt{12} = -2.165 \times 10^{-16} \quad (8)$$

Where $g$ is the value of the statistic, and $n$ is the number of samples for the two proposed models. Step 3: Establish the acceptance range for $H_o$ as $\{t: t < T_{(\alpha; n-1)}\}$ at 5% significance $\alpha = 0.05$) and $n - 1$ degrees of freedom. For the specific case, the value of $T(0.05; 11) = 1.7959$, defining the acceptance range of $H_o$ as $(-\infty, 1.7959)$ Upon evaluating the value of the statistic gg, it is observed to fall within the acceptance interval, which means that $H_o$ is not rejected. In light of the above, it can be concluded that the Max-Min Fairness algorithm can be considered as a suitable alternative for conducting optimization processes of verygood quality. This is considering the treatment comparison process with an optimization process over PL for the proposed scenario, with 95% confidence.

## 3. CONCLUSSIONS

Implementation of Max-Min Fairness as a strategy for resource optimization in IBOC FM presents clear benefits compared to current conditions. By prioritizing equity in resource allocation, this strategy offers a significant improvement in signal quality for all stations, ensuring optimal performance even for those with the worst transmission conditions. Furthermore, by reducing interferences between signals and maximizing spectrum utilization, it promotes harmonious coexistence between digital and analog transmissions, enhancing the overall quality of the radio frequency spectrum. Based on the obtained results, it was evident that the use of the MMF algorithm yielded excellent outcomes in terms of fair resource allocation with values appropriate for each service class, with 95% confidence, compared to a Linear Programming (LP) supported optimization model. Additionally, it was observed that the total bandwidth assigned to each node corresponds to the total bandwidth available in the RF channel, even under a channel saturation state.

The added value of using Max-Min Fairness extends to future research in the field of digital IBOC on FM radio. The strategy provides a solid foundation for the development of more efficient and equitable dynamic resource allocation algorithms and systems. These systems could adapt to changing channel conditions and provide optimal dynamic spectrum management, allowing for greater adaptability and flexibility in dynamic radio environments. Additionally, the Max-Min Fairness approach represents an opportunity for future research aimed at improving user experience and spectral efficiency. Furthermore, this equitable approach could be applied in more complex scenarios, including the management of multiple frequency bands and the coordination of digital radio systems in densely populated urban environments, thus opening doors to more advanced and comprehensive solutions in the field of digital broadcasting on FM.

### ACKNOWLEDGMENT

The authors of this paper want to acknowledge Minciencias and CRC through "Convocatoria 908-2021- nuevo conocimiento, desarrollotecnológico e innovación para elfortalecimiento de lossectores de TIC, postal y de contenidosaudiovisuales" for technical and financial support of the project 80740-032-2022.